\documentclass[a4paper]{jpconf}
\usepackage{graphicx}

\begin{document}

\title{Threshold Effects in Slepton Pair Production\\at the LHC}

\author{Giuseppe Bozzi}

\address{Institut f\"ur Theoretische Physik, Universit\"at Karlsruhe, P.O.Box 6980, 76128 Karlsruhe, Germany}

\ead{giuseppe@particle.uni-karlsruhe.de}

\begin{abstract}
We present a study of threshold resummation effects for slepton pair production at the Large Hadron Collider (LHC). After confirming the known NLO QCD corrections and generalizing the NLO SUSY-QCD corrections to the case of mixing squarks in the virtual loop contributions, we employ the Mellin $N$-space resummation formalism to compute logarithmically enhanced soft-gluon terms to all perturbative orders.
\end{abstract}

\section{Introduction}\label{intro}

Scalar leptons are among the lightest supersymmetric particles in many
SUSY-breaking scenarios \cite{Aguilar-Saavedra:2005pw}. They mainly decay into their Standard Model (SM) leptonic partners and the lightest supersymmetric particle. Searches for sleptons at hadron colliders will thus mainly be focused on highly-energetic lepton pairs plus missing energy. A precise prediction of the transverse-momentum spectrum of the slepton pair \cite{Bozzi:2006fw} allows to use the Cambridge (s)transverse mass to measure the slepton masses \cite{Lester:1999tx} and spin \cite{Barr:2005dz} and to extract the signal from $WW$ and $t \bar t$ production events \cite{Lytken:22,Andreev:2004qq}, which are the main backgrounds to Drell-Yan slepton pair production at the LHC. The (LO) cross section for the production of a non-mixing slepton pair was computed in \cite{Dawson:1983fw,Chiappetta:1985ku,delAguila:1990yw,%
Baer:1993ew}, while recently the mixing between the interaction eigenstates has been included \cite{Bozzi:2004qq}. The next-to-leading order (NLO) QCD
corrections have been calculated in \cite{Baer:1997nh}, and the full
SUSY-QCD corrections have been added in \cite{Beenakker:1999xh}. The genuine SUSY corrections turned out to be quite small compared to the standard QCD ones due to the presence of massive non-mixing squarks and gluino propagators in the loop diagrams.

The aim of our work \cite{Bozzi:2007qr} was to extend the previous calculations by including mixing effects relevant for the squarks appearing in the loops, and also considering the threshold-enhanced contributions due to soft-gluon emission from the initial state. These enhancements arise when the available partonic energy is just enough to produce the final state particles and thus there is a mismatch between virtual corrections and phase-space suppressed real-gluon emission. This causes the appearance of large logarithmic terms
$\alpha_s^n[\ln^{2n-1}(1-z)/(1-z)]_+$ at the $n^{{\rm th}}$ order of
perturbation theory, where $z=M^2/s$, $M$ is the slepton-pair invariant
mass, and $s$ is the partonic center-of-mass energy. Clearly, when $s$ is close to $M^{2}$, the convergence of the perturbative result is spoiled and the large logarithms have to be resummed, i.e.\ taken into account to all orders in $\alpha_{s}$. Most importantly, the convolution of the partonic cross section with the steeply falling parton distributions enhances the threshold contributions even {\it far} from hadronic threshold, i.e. when $\tau=M^2/S \ll 1$, where $S$ is the hadronic center-of-mass energy. Large corrections are thus expected for the Drell-Yan production of a slepton pair with invariant mass $M$ of a few hundreds GeV at the LHC.

The resummation of the large logarithmic contributions proceeds through the exponentiation of the soft-gluon radiation, which does not take place in $z$-space directly, but in Mellin $N$-space, where $N$ is the Mellin-variable conjugate to $z$: the threshold region $z\rightarrow 1$ corresponds to the limit $N\rightarrow \infty$. A final inverse Mellin-transform is thus required to go back to the usual $z$-space. Threshold resummation for the Drell-Yan process was first performed in \cite{Sterman:1986aj,Catani:1989ne} at the leading-logarithmic (LL) and next-to-leading-logarithmic (NLL) levels, corresponding to terms of the form $\alpha_{s}^{n}\ln^{n+1}N$ and $\alpha_{s}^{n}\ln^{n}N$ in the exponent. The extension to the NNLL level ($\alpha_{s}^{n}\ln^{n-1}N$ terms) has been carried out both for the Drell-Yan process \cite{Vogt:2000ci} and for Higgs-boson production
\cite{Catani:2003zt}. Very recently, even the NNNLL contributions
($\alpha_{s}^{n}\ln^{n-2}N$ terms) became available \cite{Moch:2005ba,Moch:2005ky,Laenen:2005uz}.

A suitable matching procedure has eventually to be performed in order to keep the full information contained in the fixed-order calculation and to apply the resummation technique only where it is fully justified. A correct matching is achieved by adding the resummed and fixed-order contributions and then subtracting the expansion of the resummed result at the same perturbative order of the fixed-order calculation: in this way, a possible double-counting of the logarithmically-enhanced contributions is avoided and a uniform theoretical accuracy over the entire invariant mass range is obtained.

\section{Numerical results}\label{numres}

Since the total cross section for slepton pair production is currently available at NLO, we could only perform soft-gluon resummation at the NLL+NLO level \cite{Bozzi:2007qr}.

We used the computer program SUSPECT \cite{Djouadi:2002ze} to calculate the physical masses of the SUSY particles and the mixing angles, and we chose the mSUGRA point SPS 1a and GMSB point SPS 7 \cite{Allanach:2002nj}, as benchmarks for our numerical studies. In the case of the lightest stau mass eigenstate $\tilde\tau_1$, which we will examine in the following, the returned value for the mass is $m_{\tilde\tau_1}$=136.2 GeV for SPS 1a and $m_{\tilde\tau_1}$=114.8 GeV for SPS 7. Feasibility studies of tau-slepton identification at the LHC with the ATLAS detector \cite{Hinchliffe:2002se} and tau tagging with the CMS detector \cite{Gennai:2002qq} have recently shown that stau masses should be observable up to the TeV range. The cross sections have been calculated both for the Tevatron, currently operating at $\sqrt S$=1.96 TeV and for the LHC, bound to operate at $\sqrt S$=14 TeV. For LO (NLO and NLL) predictions, we used the LO 2001 \cite{Martin:2002dr} (NLO 2004 \cite{Martin:2004ir}) MRST-sets of parton distribution functions.

\begin{figure}
\centering
\includegraphics[scale=0.48]{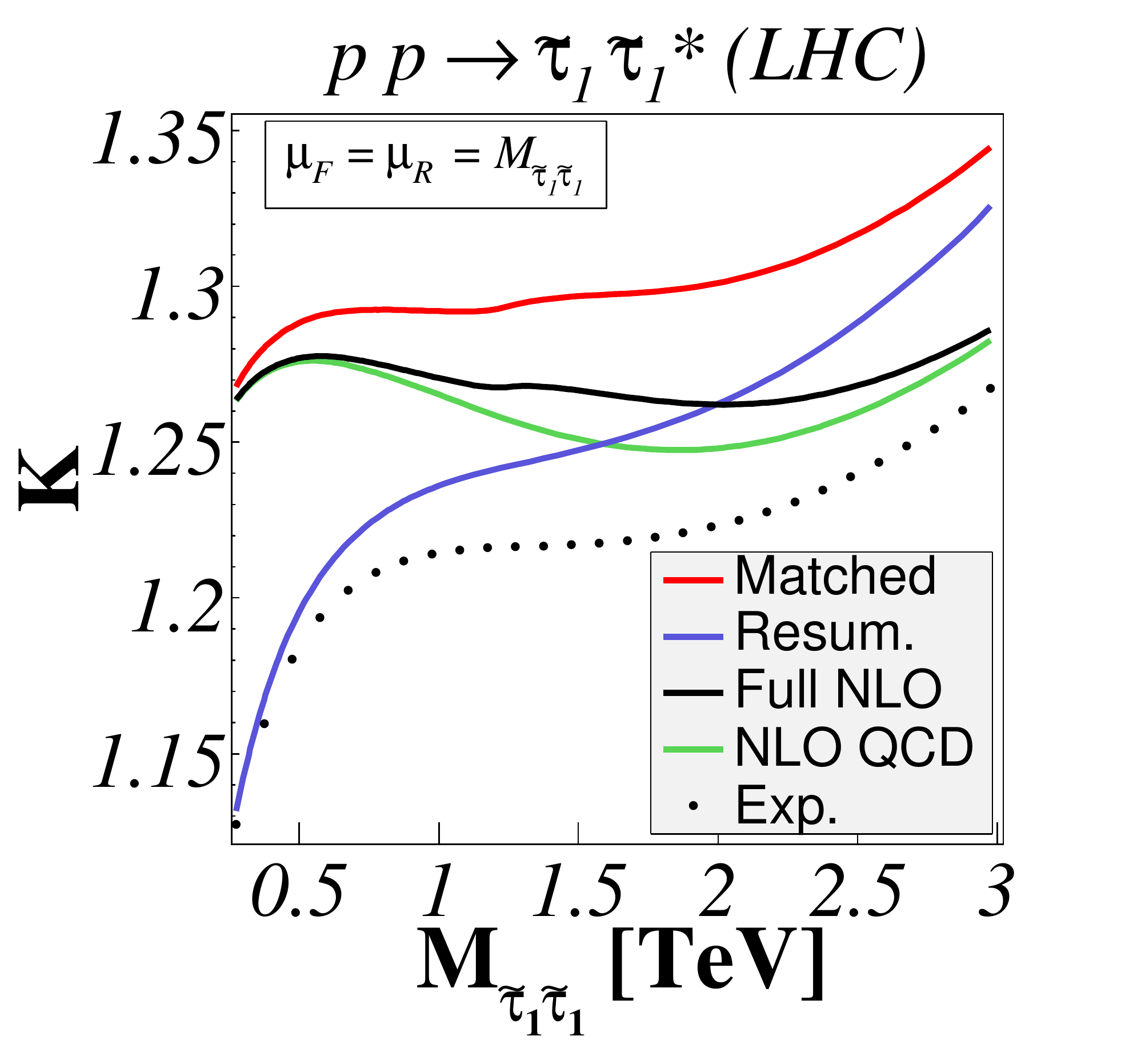}
\caption{$K$-factors for third-generation slepton-pair production at the LHC. We show the total NLL+NLO matched, the NLL resummed, the fixed order NLO (SUSY-)QCD and the expanded NLL resummed curves.}
\label{fig1}
\end{figure}

In Fig.\ref{fig1} we show the $K$-factor, with respect to the LO result, of the invariant-mass $M_{\tilde\tau_1\tilde\tau_1^*}$ distribution for stau pair production at the LHC: the total NLL+NLO matched, the NLL resummed, the fixed order NLO (SUSY-)QCD and the expanded NLL resummed curves are plotted. 

The resummed contribution mildly grows with $M$, reaching a 7\% increase over the fixed-order result for $M$=3 TeV. In this large-$M$ region, the resummed result approaches the total prediction, since the NLO QCD calculation is dominated by large logarithms and thus approaches the expanded resummed result. However, we are still far from the hadronic threshold region, so that both resummed and fixed-order contributions and a consistent matching of the two are needed. At lower values of $M$, where finite terms dominate, the resummed contribution is close to its fixed-order expansion and disappears with $M$.

The dependence on the factorization and renormalization scales has also been investigated both for the total cross section and for the invariant mass differential distribution. Within the conventional scale variations $m_{\tilde\tau_1}/2<\mu_F=\mu_R<2m_{\tilde\tau_1}$, the uncertainty reduces from 20\% at NLO to roughly 10\% after the inclusion of threshold effects.\\
{\bf Acknowledgements.} This work was partially supported by the Deutsche Forschungsgemeinschaft under SFB TR-9 ``Computergest\"utzte Theoretische Teilchenphysik''.

\end{document}